\documentclass[tightenlines,superscriptaddress,eqsecnum,floats,nofootinbib,showpacs]
{revtex4-1}
\usepackage{amsmath,amsthm,latexsym,amssymb,amsfonts,epsfig}
\usepackage{fontenc,dsfont,layout}
\usepackage{hyperref}
\usepackage{psfrag}
\usepackage[cp1250]{inputenc}
\usepackage{graphicx}
\usepackage{colordvi}
\usepackage[active]{srcltx}
\usepackage[usenames,dvipsnames]{xcolor}

\def\be{\begin{equation}}
\def\ee{\end{equation}}
\def\ba{\begin{eqnarray}}
\def\ea{\end{eqnarray}}
\def\nn{\nonumber}

\newcommand{\mubar}{{\bar \mu}} 
\newcommand{\abs}[1]{{\left|{#1}\right|}} 
\newcommand{\ket}[1]{\vert{#1}\rangle} 
\newcommand{\bra}[1]{\langle{#1}\vert} 
\newcommand{\R}{\mathcal {R}} 

\newcommand{\sgn}{\mathrm{sgn}} 
\newcommand{\grav}{\mathrm{gr}} 
\newcommand{\kin}{\mathrm{kin}} 

\begin{document}

\title{A comparison between Jordan and Einstein frames of Brans-Dicke gravity a la  loop quantum cosmology}

\author{Micha{\l} Artymowski\footnote{artymowski@bnu.edu.cn}}
\affiliation{Department of Physics, Beijing Normal University,
Beijing 100875, China}

\author{Yongge Ma\footnote{Corresponding author: mayg@bnu.edu.cn}}
\affiliation{Department of Physics, Beijing Normal University,
Beijing 100875, China}

\author{Xiangdong Zhang\footnote{zhangxiangdong@mail.bnu.edu.cn}}
\affiliation{Department of Physics, South China University of
Technology, GuangZhou 510641, China} \affiliation{Department of
Physics, Beijing Normal University, Beijing 100875, China}

\begin{abstract}
It is well known that the Jordan and Einstein frames are equivalent to each other in classical Brans-Dicke theory, provided that one and the same metric is employed for the physical space-time. Nevertheless, it is shown in this paper by cosmological models that the loop quantisation in the two different frames will lead to inequivalent effective theories. Analytical solutions have been found in both frames for the effective loop quantum Brans-Dicke cosmology without potential in: (i) vacuum case, (ii) additional massless scalar field case. In Einstein frame the analytical solution for the Brans-Dicke potential $\propto\varphi^2$ has been found. In all of those solutions the bouncing evolution of the scale factor has been obtained around Planck regime. The differences between the loop quantisation of the two frames are reflected by: (i) the evolution of the scale factor around the bounce, (ii) the scale of the bounce in the physical Jordan frame.
\end{abstract}

\maketitle

\section{Introduction}

The Brans-Dicke theory of gravity \cite{Brans:1961sx} is one of the most popular modified gravity theory, which has been investigated on various aspects for over 50 years, especial in recent decade. The scalar-tensor structure of this theory has been considered as a source of inflation and primordial inhomogeneities of space-time \cite{Starobinsky:1994mh}, dark energy \cite{Tsujikawa:2010zza,Qiang,Qiang:2009fu} and in the context of stability of stars \cite{Babichev:2009fi}. The newest results on the Cosmic Microwave Background \cite{Ade:2013uln} shows that the $R^2$ inflation, which may be also expressed in terms of Brans-Dicke field \footnote{In rest parts of this paper we will refer $f(R)$ theories to their Brans-Dicke form in metric formalism.}, fits the data of the spectrum of perturbations. On the other hand, as the background independent quantisation of general relativity (GR), loop quantum gravity (LQG) \cite{Ashtekar:2004eh,Han:2005km} has been rather active in recent two decades. The expectation that the singularity predicted by classical GR would be resolved by quantum gravity has been confirmed by the recent study of the loop quantum cosmology (LQC)  \cite{Bojowald:2001xe,Ashtekar:2006wn}, which is a simplified, symmetric model of LQG \cite{Bojowald:2006da,Ashtekar:2011ni}. The Big-Bang singularity in the cosmological model of GR is replaced by the quantum bounce of LQC. Recently, the nonperturbative quantisation scheme of LQG has been successfully extended to $f(R)$ theories \cite{Zh11, Zh11b} and Brans-Dicke theory \cite{Zhang:2011gn}-\cite{ZAM}. The corresponding cosmological model for Brans-Dicke theory has been set up \cite{ZAM}. The purpose of this paper is to compare the Jordan frame with Einstein frame of Brans-Dicke gravity by their loop quantum cosmology models. Note that the original formulation of Brans-Dicke theory was in Jordan frame. If one and the same metric is employed to represent physical space-time, the Jordan and Einstein frames are equivalent to each other in classical Brans-Dicke theory. However, there is no guarantee for the equivalence of the quantisation in two frames. As shown in \cite{Zhang:2011vg} the quantisation procedure of Brans-Dicke theory shall distinguish between two cases: $\omega = -\frac{3}{2}$ and $\omega \neq -\frac{3}{2}$. In this paper we shall assume that $\omega \neq -\frac{3}{2}$. This assumption comes from observational limitations on Brans-Dicke theory, which prefers $\omega \gg 1$ \cite{will,will1}.
\\*

To transform the Brans-Dicke theory into Einstein frame one has to redefine the metric tensor, which would cause the canonical form of the GR action. The LQC in Einstein frame has been studied in Refs.\cite{Artymowski:2012is,Bojowald:2006bz,Bojowald:2006hd}. In this paper we take the original idea of Brans and Dicke that Jordan frame is the physical one \footnote{For arguments see \cite{Barvinsky:2008ia}.}, though in general this remains open. For instance the quantisation in different frames may give different results of the evolution of primordial gravitational waves, which in the future could help us to discriminate one frame and favour the other. Following interpretation of Jordan frame as the physical one we compare two methods of LQC quantisation: in Jordan and Einstein frame. In latter case we shall transform results into Jordan frame for precise comparison.
\\*

In this paper, according to Refs.\cite{ACS,ZAM} and for simplicity consideration, we only focus on the effective LQC of the two frames, where holonomy corrections are included, while neglecting inverse triad corrections. Therefore, by the LQC correction we will mean LQC holonomy corrections to the flat FRW space-time in different frames. This treatment is usually considered to be realistic and consistence with effective equations of LQC of GR. All calculations in this paper are performed in Planck units, i.e. for $8\pi G = M_{pl}^{-2} = 1$, {$\hbar=1$}.
\\*

The structure of this paper goes as follows: In Sec. \ref{sec:class} we introduce classical Hamiltonian of Brans-Dicke theory in Jordan frame in flat FRW model. In Sec: \ref{sec:LQCJordan} we calculate effective equations of motion in semi-classical approach to LQC Brans-Dicke theory in Jordan frame. In Sec. \ref{sec:exact} we present exact solutions of semi-classical equations of motion for LQC Jordan frame quantisation for the vacuum case and for the additional massless scalar field case. In Sec. \ref{sec:einstein} we introduce the Hamiltonian formalism of the same cosmological model of Brans-Dicke theory in Einstein frame and its semi-classical equations of motion for LQC Einstein frame quantisation. In Sec. \ref{sec:einsteinexact} we solve the semi-classical equations for Einstein frame analytically and compare the results of the LQC quantisation in both frames. Finally we conclude in Sec. \ref{sec:concl}.

\section{Classical Brans-Dicke theory} \label{sec:class}

We start with the classical Brans-Dicke theory coupled with a scalar matter field. The Jordan frame action reads
\ba
S(g,\varphi,\chi)=\frac{1}{2}\int_\Sigma
d^4x\sqrt{-g}\left[\varphi\R-\frac{\omega}{\varphi}(\partial_\mu\varphi)\partial^\mu\varphi-2V(\varphi)-(\partial_\mu\chi)\partial^\mu\chi-2W(\chi)\right]\ ,\label{action}
\ea
where $\varphi$ is the Brans-Dicke scalar field, $\chi$ is a scalar matter field,  $V(\varphi)$ and $W(\chi)$ are potentials respectively. Now we consider an isotropic and homogenous $k=0$ Universe. We choose a fiducial
Euclidean metric  $ {}^oq_{ab}$ on the spatial slice of the isotropic observers and introduce a pair of fiducial orthonormal
triad and co-triad as $({}^oe^a_i , {}^o\omega^i_a)$ respectively
such that $ {}^oq_{ab}={}^o\omega^i_a{}^o\omega^i_b$.
Then the physical spatial metric is related to the fiducial by $ q_{ab}=a^2 {}^oq_{ab}$, and its line element can be described by the Friedman-Robertson-Walker (FRW) form
\ba
ds^2=-dt^2+a^2(t)\left(dr^2+r^2(d\theta^2+\sin^2\theta d\phi^2)\right) \nn
\ea
where $a$ is the scale factor. Then the classical action (\ref{action}) reduces to
\begin{equation}
\mathcal{L}=-3\dot{a}^2a\varphi - 3\dot{a}a^2\dot{\varphi} +a^3\frac{\omega }{2}\frac{\dot{\varphi}^2}{\varphi } - a^3V(\varphi)+a^3\frac{1}{2}\dot{\chi}^2 - a^3W(\chi)\ ,
\end{equation}
{By the Legendre transformation the} canonical momenta {read respectively} as
\begin{equation}
\pi_\varphi = \frac{\partial\mathcal{L}}{\partial\dot{\varphi}} = a^2 \left(-3 \dot{a}+\frac{\omega  a\dot{\varphi}}{\varphi}\right)\ ,\qquad \pi_\chi = \frac{\partial\mathcal{L}}{\partial\dot{\chi}} = a^3\dot{\chi} \ ,\qquad \pi_a = \frac{\partial\mathcal{L}}{\partial\dot{a}} = -3a\left(2\varphi\dot{a}+a\dot{\varphi}\right) \ .
\end{equation}
Therefore, from $\mathcal{H} = \pi_\varphi\dot{\varphi}+\pi_\chi\dot{\chi}+\pi_a\dot{a}-\mathcal{L}$ one obtains the classical Hamiltonian as a function of $\varphi$, $\chi$, $a$ and their canonical momenta {as}
\begin{equation}
\mathcal{H}_{class}(a,\pi_a)=-\frac{1}{6\beta a^3}\left(-3\beta \left(\pi_\chi^2 + 2a^6 (V(\varphi )+W(\chi ))\right) - 6\varphi\pi_\varphi^2 + 6a\pi_\varphi\pi _a + \frac{\omega}{\varphi}a^2\pi_a^2\right)\ ,
\end{equation}
where $\beta := 2\omega + 3$. {While the spatial slice of our cosmological model is infinite, we may introduce an ``elemental
cell" $\mathcal {V}$ and restrict all
integral to $\mathcal {V}$. For simplicity,
we let the elemental cell $\mathcal {V}$ be a cubic measured by our
fiducial metric and denotes its volume as $V_o$. Via fixing the degrees of freedom of local
gauge and diffeomorphism transformations, we finally obtain the connection and
densitized triad by symmetrical reduction as \cite{LQC5}:
\ba
A_a^i=c
V_0^{-\frac13}{}^o\omega^i_a,\quad\quad\quad
E^b_j=pV_0^{-\frac23}\sqrt{\det({}^0q)}{}^oe^b_j\ , \label{eq:canonicalvariables}
\ea
where $c,p$
are only functions of $t$.} {Note that the new variables are related to the old ones by
\begin{equation}
|p|=a^2V_0^{\frac 23}\ ,\qquad {c =- \gamma\sgn(p)\frac{\pi _a}{6a}V_0^{\frac 13}}\ ,\label{eq:ashtekar}
\end{equation}}
where $\gamma$ is the so-called Barbero-Immirzi parameter. {Now the gravitational part of the phase space of the cosmological model consists of conjugate pairs $(c,p)$ and
$(\varphi,\pi_\varphi)$. The basic Poisson brackets between them can be simply
read as\ba
\{c,p\}&=&\frac{1}{3}\gamma,\nn\\
\{\varphi,\pi_\varphi\}&=&1. \label{poissonb}\ea}
Thus, the classical Hamiltonian in terms of {new} variables is of the form{
\begin{equation}
\mathcal{H}(c,p)=\frac{-6\frac{\omega}{\varphi}c^2p^2 + 6\gamma cp\pi_\varphi + \gamma^2\left(\frac{\beta}{2}\pi_\chi^2+\beta  |p|^3 (V(\varphi )+W(\chi ))+\varphi\pi_\varphi ^2\right)}{\gamma^2\beta |p|^{3/2}}\ .\label{eq:Hamcp}
\end{equation}}
From the Hamiltonian equations one obtains following classical equations of motion.
\begin{equation}
\dot{\xi}=\{\xi,\mathcal{H}\}=\frac{\gamma}{3} \left( \frac{\partial\xi}{\partial c}\frac{\partial \mathcal{H}}{\partial p} - \frac{\partial\xi}{\partial p}\frac{\partial \mathcal{H}}{\partial c} \right) + \frac{\partial\xi}{\partial \varphi}\frac{\partial \mathcal{H}}{\partial \pi_\varphi} - \frac{\partial\xi}{\partial \pi_\varphi}\frac{\partial \mathcal{H}}{\partial \varphi} \ , \label{eq:hameq}
\end{equation}
where $\xi=\xi(c,p,\varphi,\pi_\varphi)$ is some function on the classical phase space.
\\*

Let us generalize {above equations to} the Brans-Dicke theory {coupled} with any perfect fluid. Then {combining} the Hamiltonian equations and the scalar constraint $\mathcal{H}=0$ one obtains
\begin{eqnarray}
\ddot{\varphi} + 3H\dot{\varphi} + \frac{2}{\beta}(\varphi V_\varphi - 2V(\varphi)) &=& \frac{1}{\beta}\left(\rho_M - 3P_M\right)\ ,\label{eq:motionBD}\\
3\left(H + \frac{\dot{\varphi}}{2\varphi}\right)^2 = \frac{\beta}{4}\left(\frac{\dot{\varphi}}{\varphi}\right)^2 +\frac{V(\varphi)}{\varphi} + \frac{\rho_M}{\varphi} &=& \frac{1}{\varphi^2}\rho_e\ , \label{eq:FriedBD}
\end{eqnarray}
where $V_\varphi \equiv \frac{dV}{d\varphi}$, {$H:=\frac{\dot{a}}{a}$ is the Hubble parameter, and} {$\rho_e := \frac{\beta}{4}\dot{\varphi}^2+\varphi(V(\varphi)+\rho_M)$} is the effective energy density. For $V=\rho_M = P_M = 0$ equations (\ref{eq:motionBD}) {and} (\ref{eq:FriedBD}) have analytical solution of the form
\begin{equation}
H=\frac{\dot{\varphi}}{2\varphi}\left(\pm\sqrt{\frac{\beta}{3}}-1\right)\ ,\qquad \dot{\varphi} = \frac{1}{2} \sqrt{\beta \rho _i} \left(\frac{\varphi }{\varphi _i}\right)^{\pm \frac{1}{2} \left(3-\sqrt{3\beta }\right)}\ ,
\end{equation}
where $\varphi_i$ is {some initial value of $\varphi$}. For $\beta=3$ (i.e., $\omega=0$) one obtains
\begin{equation}
H=0\ ,\quad\varphi=const\qquad \lor \qquad H = \frac{1}{2t}\ ,\quad\varphi = \sqrt{\frac{2\alpha_1}{\alpha_2-t}}\ .\label{eq:VacuumSolution}
\end{equation}

\section{LQC corrections to Brans-Dicke theory in Jordan frame}\label{sec:LQCJordan}

{Let us} {consider LQC corrections to the Brans-Dicke theory we mentioned above.} {In this paper we follow the hybrid approach: the connection and triad are quantised by the polymer-like quantisation, while all other canonical variables are quantised by the Schrodinger quantisation.} {The kinematic Hilbert space for the geometry part can be defined as
$\mathcal{H}_{\kin}^{\grav}:=L^2(R_{Bohr},d\mu_{H})$, where
$R_{Bohr}$ and $d\mu_{H}$ are respectively the Bohr
compactification of the real line (the configuration space) and Haar measure on it
\cite{LQC5}, while the kinematic Hilbert spaces for the scalar fields are defined as
in usual quantum mechanics. The whole Hilbert
space is their direct product. Let $\ket{\mu}$ be the eigenstates of
 $\hat{p}$ in the kinematic Hilbert space $\mathcal{H}_{\kin}^{\grav}$ such that
 \ba
\hat{p}\ket{\mu}=\frac{\gamma}{6}\mu\ket{\mu}\ .\label{eq:ponmu}
\ea
It turns out that those
states satisfy the following orthonormal condition
\ba
\bra{\mu_i}{\mu_j}\rangle=\delta_{\mu_i,\mu_j}\ , \ea
where $\delta_{\mu_i,\mu_j}$ is the Kronecker delta function rather than the Dirac distribution. Note that in LQC framework, while there is no operator corresponding the connection $c$, its holonomy $\exp(i\mu c/2)$ along a line with oriented length $\mu$ is a well-defined operator. In the improved dynamics setting \cite{Ashtekar:2006wn}, one employs the length $\mubar=\sqrt{\frac{\Delta}{\abs{p}}}$, with
$\Delta=4\sqrt{3}\pi\gamma{\ell}_{\textrm{p}}^2$ being a minimum
nonzero eigenvalue of the area operator \cite{Ash-view,Ashtekar:1996eg}, to construct the Hamiltonian constraint operator. In the semiclassical regime, as a basic variable, the holonomy will certainly lead to corrections to the classical equations.} {Here we only focus on the LQC holonomy correction, while neglecting inverse triad corrections}. {A heuristic and simple way to
get the holonomy corrections is to replace the connection variable by its holonomy, i.e., $c\to\frac{\sin(\mubar c)}{\mubar}$, though its validity should be checked by detailed calculations.}

\subsection*{Friedmann equation}

{By the following substitution:}
\begin{equation}
\varphi \to \varphi\ ,\quad \pi_\varphi \to \pi_\varphi\ ,\quad \chi \to \chi\ , \quad\pi_\chi \to \pi_\chi\ ,\quad p \to p\ ,\quad c \to \sqrt{\frac{|p|}{\Delta}}\sin\left(c\sqrt{\frac{\Delta}{|p|}}\right)\ ,\label{eq:loop}
\end{equation}
{an effective Hamiltonian constraint with holonomy corrections of loop quantum Brans-Dicke cosmology can be obtained from Eq. (\ref{eq:Hamcp}) as}
{\begin{eqnarray}
\mathcal{H}_{LQC}=\frac{1}{|p|^{3/2} \beta  \gamma  \Delta }\left[ \pi _{\varphi } \left(6 p|p|^{1/2} \sqrt{\Delta } \sin\left(c \sqrt{\frac{\Delta }{|p|}}\right)+\gamma  \Delta  \varphi  \pi _{\varphi }\right)+\frac{\beta }{2} \gamma  \Delta  \pi _{\chi }^2+\right.\nonumber\\
 |p|^3 \left(-6 \frac{\omega }{\gamma  \varphi } \sin^2\left(c \sqrt{\frac{\Delta }{|p|}}\right)+\beta  \gamma  \Delta  \left(V(\varphi )+W(\chi)\right)\right)\Bigg]\ .
\end{eqnarray}}
{Note that this effective Hamiltonian constraint can also be derived by a systematic approach as in Ref.\cite{ZAM}.}
All {semi-classical} equations of motion {can} be obtained from the {Eq.} (\ref{eq:hameq}) with $\mathcal{H}_{LQC}$ as a Hamiltonian. For instance, from $\dot{\varphi}=\{\varphi,\mathcal{H}_{LQC}\}$ one obtains{
\begin{equation}
\dot{\varphi} = \frac{6p\sin\left(c \sqrt{\frac{\Delta }{|p|}}\right)}{\beta \sqrt{\Delta }\gamma|p| }+\frac{2 \varphi\pi _{\varphi}}{\beta  |p|^{3/2}}\quad \Rightarrow\quad \pi_\varphi = |p|^{3/2}\frac{ -6 \sgn(p)\sin\left(c \sqrt{\frac{\Delta }{|p|}}\right)+\gamma  \sqrt{\Delta } \beta \dot{\varphi}}{2 \gamma  \sqrt{\Delta } \varphi }\ . \label{eq:pivarphi}
\end{equation}}
From the equation $\dot{p} = 2pH = 2p\frac{\dot{a}}{a} = \{p,\mathcal{H}_{LQC}\}$, one finds{
\begin{equation}
H = \frac{\cos\left(c \sqrt{\frac{\Delta }{|p|}}\right) \left(2 \omega  p|p|^{1/2} \sin\left(c \sqrt{\frac{\Delta }{|p|}}\right)-\gamma  \sqrt{\Delta }\varphi  \pi _{\varphi }\right)}{\gamma  \beta  \sqrt{\Delta } |p|^{3/2} \varphi } \ ,\label{eq:Hubble}
\end{equation}}
where $H$ is a Hubble parameter. Substituting Eq. (\ref{eq:pivarphi}) into the scalar constraint $\mathcal{H}_{LQC}=0$ one obtains{
\begin{equation}
\sin^2\left(c \sqrt{\frac{\Delta }{|p|}}\right) = \frac{\gamma^2\Delta}{3}\left(\frac{\beta}{4}\dot{\varphi}^2 + \varphi\frac{1}{2}\dot{\chi}^2 +\varphi V(\varphi ) + \varphi W(\chi )\right)\ .\label{eq:sin}
\end{equation}}
From Eqs. (\ref{eq:pivarphi}),(\ref{eq:Hubble}),(\ref{eq:sin}) and $\mathcal{H}_{LQC}=0$ one finds the semi-classical LQC version of the first Friedmann equation,
\begin{equation}
\left(H+\frac{\dot{\varphi }}{2\varphi }\right)^2 = \left(\frac{1}{\varphi }\sqrt{\frac{\rho _e}{3}} \sqrt{1-\frac{\rho _e}{\rho _{\text{cr}}}}+\frac{\dot{\varphi}}{2\varphi } \left(1-\sqrt{1-\frac{\rho _e}{\rho _{\text{cr}}}}\right)\right)^2\ , \label{eq:FriedBDJordan}
\end{equation}
where $\rho_e = \frac{\beta }{4} \dot{\varphi }^2+ \varphi  \left(\frac{1}{2}\dot{\chi}^2+V(\varphi)+W(\chi)\right)$ is the effective energy density and $\rho_\text{cr} = \frac{3}{\gamma^2\Delta}\simeq 0.41G^{-2}\simeq 260 M_{pl}^4$ is the critical (maximal) energy density. {Eq (\ref{eq:FriedBDJordan}) coincides with the effective Friedmann equation in Ref. \cite{ZAM}, where the potentials of scalar fields are not included.}

\subsection*{Equation of motion of $\varphi$}

From Eq. (\ref{eq:pivarphi}) we define{
\begin{equation}
\tilde{p}_\varphi = |p|^{3/2}\dot{\varphi} = \frac{6p|p|^{1/2} \sin\left(c \sqrt{\frac{\Delta }{|p|}}\right)}{\beta \sqrt{\Delta }\gamma }+\frac{2}{\beta }\varphi  \pi _{\varphi }\ .
\end{equation}}
{Then we have}
\begin{equation}
{\dot{\tilde{p}}_\varphi = \frac{d}{dt}(a^3\dot{\varphi}) = a^3(\ddot{\varphi}+3H\dot{\varphi}) = \{\tilde{p}_\varphi,\mathcal{H}_{LQC}\}}\ .\label{eq:tildep}
\end{equation}
{From Eq.} (\ref{eq:pivarphi}),(\ref{eq:sin}) and (\ref{eq:tildep}) one obtains
\begin{equation}
\ddot{\varphi} + 3 H\dot{\varphi } + \frac{2}{\beta}\varphi V_\varphi + \frac{2}{\beta} \left(V(\varphi) + W(\chi)\right)\left (1 - 3\sqrt {1 - \frac {\frac {\beta} {4}\dot {\varphi }^2 + \varphi\left(V(\varphi) +  W(\chi) + \frac {1}{2}\dot{\chi}^2 \right)} {\rho_{\text{cr}}}} \right) = - \frac{\dot {\chi}^2}{\beta}\ .\label{eq:EOMvarphi}
\end{equation}
This equation may be expressed in more general way by {expressing the scalar field} $\chi$ with {its} energy density $\rho_M$ and pressure $P_M$ {as}
\begin{equation}
\ddot{\varphi} + 3 H\dot{\varphi } + \frac{2}{\beta}\varphi V_\varphi + \frac{2}{\beta} \left(V(\varphi) + \rho_M - P_M\right)\left (1 - 3\sqrt {1 - \frac {\frac {\beta} {4}\dot {\varphi }^2 + \varphi\left(V(\varphi) +  \rho_M\right)} {\rho_{\text{cr}}}} \right) = - \frac{1}{\beta}(\rho_M+P_M)\ . \label{eq:EOMvarphiGeneral}
\end{equation}
{Then} the $\chi$ field could be replaced by any other perfect fluid, e.g. by dust, radiation, cosmological constant etc. From $\dot{\pi}_\chi=\{\pi_\chi,\mathcal{H}\}$, one obtains the equation of motion of $\chi$ {as}
\begin{equation}
\ddot{\chi} + 3H\dot{\chi} + W_\chi = 0\ ,
\end{equation}
{where $W_\chi\equiv \frac{dW}{d\chi}$}. The LQC corrections enters this equation {due to the existence of} potential terms of $\varphi$ and $\chi$. Even for $V(\varphi)=0$ this correction may appear due to the existence of a non-zero $W(\chi)$. Therefore, the continuity equation is not modified by LQC quantisation in Jordan frame as long as $V(\varphi) = W(\chi) = 0$, which is the case considered in Ref.\cite{Zhang:2011gn}.
For $\rho_e \ll \rho_\text{cr}$,  {from Eq.(\ref{eq:EOMvarphiGeneral})} one obtains
\begin{equation}
\ddot{\varphi} + 3 H\dot{\varphi } + \frac{2}{\beta}(\varphi V_\varphi - 2V(\varphi)) \simeq \frac{1}{\beta}(4 W(\chi) - \dot {\chi}^2) \simeq \frac{1}{\beta}(\rho_\chi - 3P_\chi)\ ,
\end{equation}
which recovers the {classical} {equation} (\ref{eq:motionBD}).

\section{Exact solutions of effective loop quantum Brans-Dicke cosmology in Jordan frame} \label{sec:exact}

\subsection*{Vacuum solution}

Let us consider vacuum solution of {effective loop quantum} Brans-Dicke {cosmology} in Jordan frame with $V(\varphi)=0$.  Under the assumption $\rho_M = P_M = 0$ one obtains
\begin{equation}
\ddot{\varphi} + 3H\dot{\varphi} = 0\ \Rightarrow\ \left(\frac{\dot{\varphi }}{2\varphi } - \frac{\ddot{\varphi }}{3\dot{\varphi }}\right)^2 =  \left(\frac{\dot{\varphi }}{2\varphi }\right)^2\left(\left(\sqrt{\frac{\beta }{3}}-1\right)\sqrt{1 - \frac{\beta }{4\rho _{\text{cr}}}\dot{\varphi }^2} + 1\right)^2 \label{pphidot}\ .
\end{equation}
Since $\dot{\varphi}\propto a^{-3}>0$, one can use $\varphi$ as a time variable. Thus, let us consider $\dot{\varphi}$ as a function of $\varphi$, i.e. $\dot{\varphi} = f(\varphi)$. This implies $\ddot{\varphi}=f\varphi f$, where $f_\varphi=\frac{df}{d\varphi}$. {We already got the analytic solution of $f$ and $H$ in Ref.\cite{ZAM}, where there exists a quantum bounce. In this paper, in order to see the singularity resolution more {explicitly} and also for a comparison with the Einstein frame quantization. We will plot the evolution of {the} volume {of the elemental cell} with respect {to the} scalar time.} From Eq. (\ref{pphidot}) one obtains the analytical formula for the scale factor as a function of $\varphi$,
\begin{equation}
\frac{1}{a}\frac{da}{dt}=\frac{a_\varphi}{a}f=f\frac{d}{d\varphi}(\log a) = -\frac{1}{3}f_\varphi\ \Rightarrow a(\varphi)\propto f^{-1/3}\ .
\end{equation}
The evolution of the volume of the {elemental cell} $V= a(\varphi)^3V_o$ in this case, as well as {its} comparison with the classical evolution is shown {in} Fig. \ref{fig:Voffields}.
\\*

The main motivation to consider the vacuum case without potential (as well as the additional massless scalar field scenario, which will be mentioned later) is {due to} the characteristic feature of theories with a quantum bounce: around the bounce kinetic terms of fields shall dominate over potentials. This means that the solutions obtained here shall also be a good approximation of the evolution around the bounce in more realistic theories with inflationary potentials.

\subsection*{Brans-Dicke cosmology with massless scalar field}

{Now we would like to extend the above results to incorporate a massless scalar field as an outside matter field.} Let us consider Eq. (\ref{eq:FriedBDJordan}) for $\rho_e=\frac{\beta}{4}\dot{\varphi}^2 + \frac{\varphi}{2}\dot{\chi}^2$, where $\chi$ is a massless scalar field. {Since} the LQC correction does not modify {the} conservation law for $V=W=0$, one obtains
\begin{equation}
\ddot{\varphi} + 3H \dot{\varphi} = -\frac{1}{\beta}\dot{\chi}^2\ ,\qquad \ddot{\chi} + 3H \dot{\chi} = 0 \ \Rightarrow\ H = -\frac{1}{3}\frac{\ddot{\chi}}{\dot{\chi}}\ . \label{eq:motionBDfield}
\end{equation}
From Eq. (\ref{eq:motionBDfield}) one finds the analytical relation between $\varphi$ and $\chi$
\begin{equation}
\varphi = -\frac{1}{2\beta} \chi^2+B \chi + C\ ,\qquad \chi \in \left(B \beta -\sqrt{\beta}\sqrt{2C+B^2\beta}\ ,\ B\beta +\sqrt{\beta}\sqrt{2C+B^2\beta}\right) \label{eq:phiofvarphi}
\end{equation}
where {$B$ and $C$ are constants satisfying} $B=\frac{\dot{\varphi}_{\text{cr}}}{\dot{\chi}_{\text{cr}}}+\frac{\chi_{\text{cr}}}{\beta}$ {and} $C=\varphi_{\text{cr}}-\frac{\chi_{\text{cr}}^2}{\beta}-\frac{\dot{\varphi}_{\text{cr}}}{\dot{\chi}_{\text{cr}}}\chi_{\text{cr}}$ {respectively. Here the subscript} $_{\text{cr}}$ denotes the value of the field at the moment of the bounce. Since $\dot{\varphi}$ may {change its sign} during the {time} evolution one can not use the Brans-Dicke field as a time variable. However, massless scalar field $\chi$ remains monotonic {with respect to the cosmological time. Hence} in following analysis we shall use it to parametrize time flow. {Note that the limit} $\chi\to B\beta \pm \sqrt{\beta}\sqrt{2C+B^2\beta}$ correspond to $t\to\pm\infty$. {The relation (\ref{eq:phiofvarphi}) can be also obtained for the Einstein frame quantisation as well as in the classical Brans-Dicke theory.}
\\*

The GR limit {of} Brans-Dicke theory is obtained for $\varphi\to 1$. In considered scenario $\chi$ is always growing with time, because $\dot{\chi}$ is always positive. Since for late time (big $\chi$) $\varphi\to 0$, one does not recovers GR limit. However, this shall not be a problem of this analysis, since the physical evolution of $\varphi$ at late time depends on the other matter fields which would fill the universe. We have assumed that around the bounce the universe is dominated by $\chi$. But this assumption is not realistic from the point of view of observable Universe. With $V=W=0$ one does not obtain inflation, reheating, bariogenesis etc. Therefore, for energy much smaller than $\rho_{\text{cr}}$ one needs to consider the existence of additional fields, or at least potential terms of $\varphi$ or $\chi$. This, due to Eq. (\ref{eq:EOMvarphiGeneral}), shall modify the evolution of $\varphi$, which could obtain the desired limit $\varphi\to 1$.
\\*

\begin{figure}[h]
\centering
\includegraphics*[height=6.45cm]{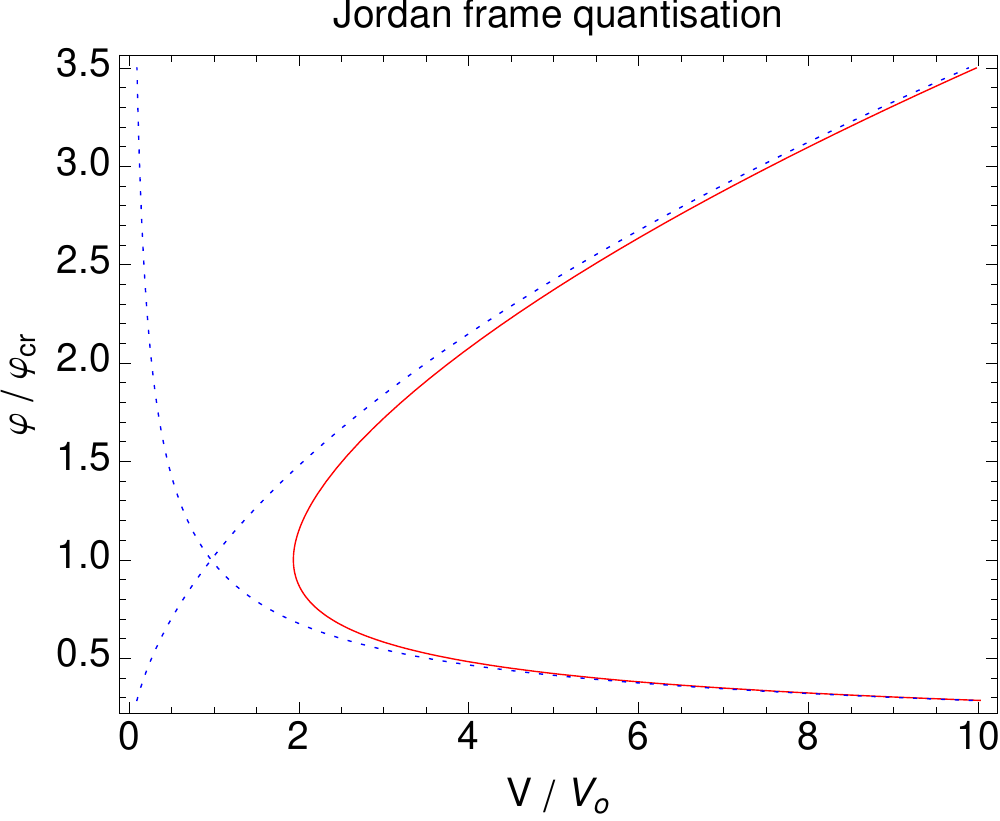}
\hspace{0.55cm}
\includegraphics*[height=6.45cm]{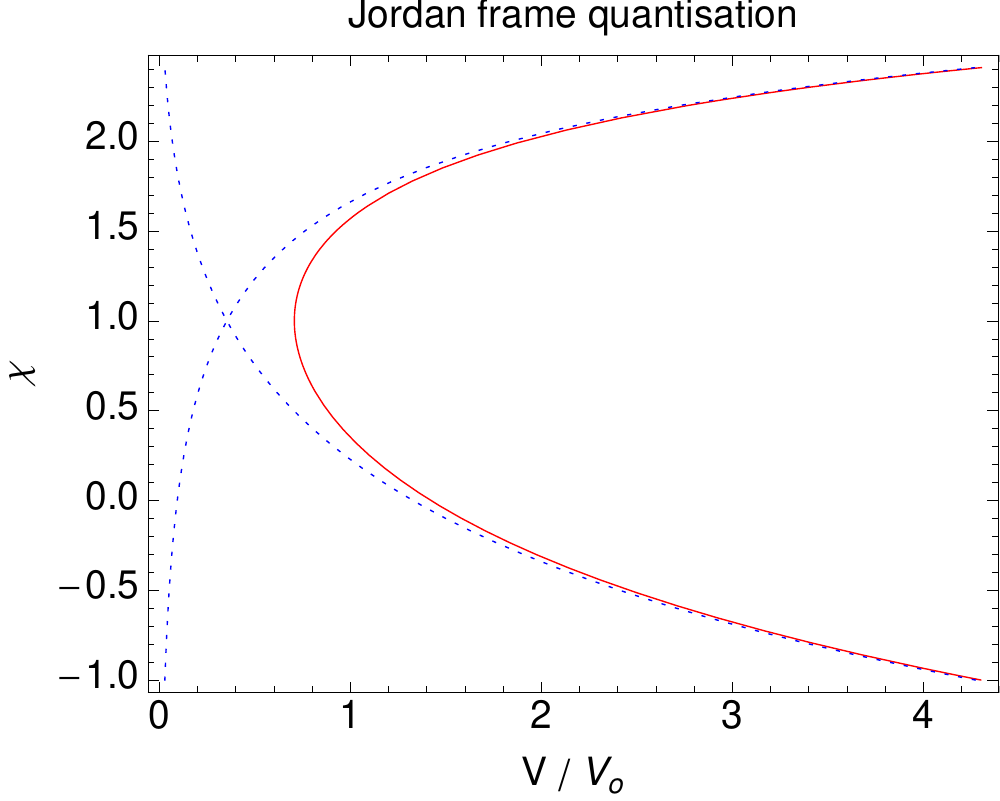}\\
\ \\
\includegraphics*[height=6.45cm]{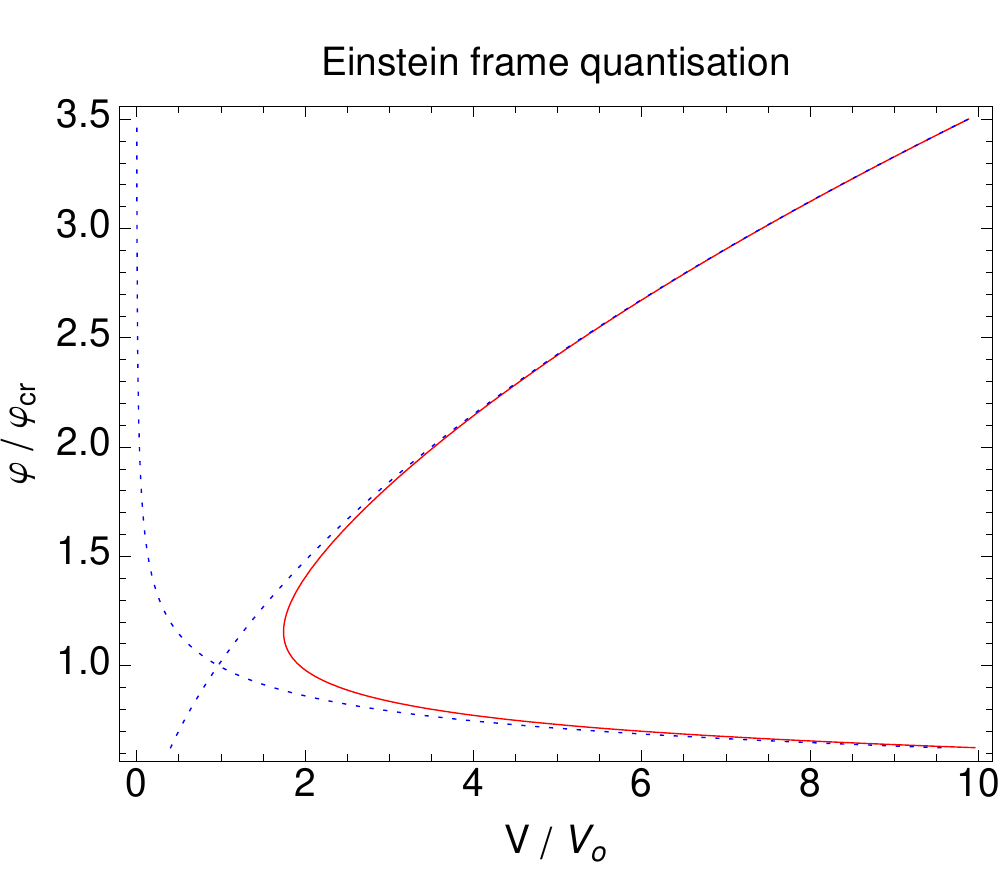}
\hspace{0.55cm}
\includegraphics*[height=6.45cm]{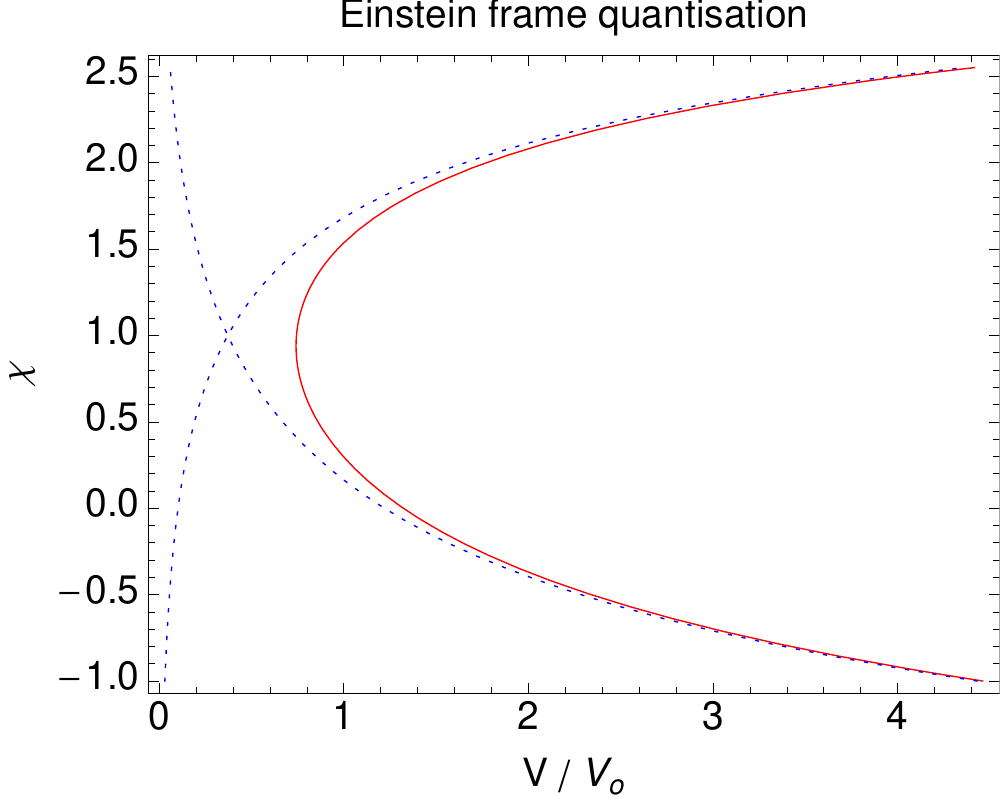}
\caption{\it All panels present analytical results for the evolution of the volume of the elemental cell of the universe as a function of $\varphi$ (vacuum case, left panels) and $\chi$ (massless scalar field case, right panels, $\chi_{\text{cr}}=1$) for various $\beta\sim O(10)$. Solid red lines represent the solution for effective LQC, while dotted blue lines represent classical solutions with singularity. Unlike the case of LQC GR, the LQC Brans-Dicke provides asymmetric evolution of $V$.}
\label{fig:Voffields}
\end{figure}

To obtain analytical solution for the Hubble parameter let us note that Eq. (\ref{eq:FriedBDJordan}) may be rewritten as a second order differential equation of the $\chi$ field {by taking account of} Eqs. (\ref{eq:FriedBDJordan}),(\ref{eq:motionBDfield}) {and} (\ref{eq:phiofvarphi}) {as}
\begin{equation}
\ddot{\chi} + \frac{\dot{\chi}^2 \sqrt{1-\frac{\left(2 C+B^2 \beta \right) \dot{\chi}^2}{4\rho_{\text{cr}} }} \left(3 (\chi -B \beta )+ \beta  \sqrt{3\left(2 C+B^2 \beta \right)}\right)}{2 C \beta +2 B \beta  \chi -\chi ^2} = 0 \ . \label{eq:varphioft}
\end{equation}
The effective energy density is now equal to
\begin{equation}
\rho_e = \frac{1}{4}\left(2C+B^2\beta\right)\dot{\chi}^2\ .
\end{equation}
This comes from Eq. (\ref{eq:phiofvarphi}) {and} the fact that $\rho_\chi \propto \rho_e\propto a^{-3}$. Since Eq. (\ref{eq:varphioft}) does not contain any explicit dependence of t, one can substitute $\dot{\chi}$ by $g(\chi)=\dot{\chi}$. Then one obtains
\begin{equation}
g_\chi+g\frac{\left( \sqrt{3\left(2 C+B^2 \beta \right) }-3 B+3\frac{\chi }{\beta }\right)}{\left(2 C+2 B \chi - \frac{\chi ^2}{\beta }\right)} \sqrt{1-\frac{\left(2 C+B^2 \beta \right) g^2}{4\rho _{\text{cr}}}} = 0\ ,
\end{equation}
where $g_\chi=\frac{dg}{d\chi}$. The exact solution of this equation is
\begin{equation}
g(\chi) = \frac{4\sqrt{\rho _{\text{cr}}}}{\sqrt{2 C+B^2 \beta }}\frac{\left(\frac{B \beta +\sqrt{\beta } \sqrt{2 C+B^2 \beta }-\chi }{B \beta +\sqrt{\beta } \sqrt{2 C+B^2 \beta }-\chi _{\text{cr}}}\right)^{ \frac{1}{2}\left(\sqrt{3\beta }+3\right)}\left(\frac{-B \beta +\sqrt{\beta } \sqrt{2 C+B^2 \beta }+\chi }{-B \beta +\sqrt{\beta } \sqrt{2 C+B^2 \beta }+\chi _{\text{cr}}}\right)^{\frac{1}{2}\left(\sqrt{3\beta }-3\right)} }{\left(\frac{B \beta +\sqrt{\beta } \sqrt{2 C+B^2 \beta }-\chi }{B \beta +\sqrt{\beta } \sqrt{2 C+B^2 \beta }-\chi _{\text{cr}}}\right)^{\sqrt{3\beta }+3}+\left(\frac{-B \beta +\sqrt{\beta } \sqrt{2 C+B^2 \beta }+\chi }{-B \beta +\sqrt{\beta } \sqrt{2 C+B^2 \beta }+\chi _{\text{cr}}}\right)^{\sqrt{3\beta }-3}}\ .
\end{equation}
The Hubble parameter is equal to $H = -\frac{1}{3}\frac{\ddot{\chi}}{\dot{\chi}} = -\frac{1}{3}g_\chi$. Analogously to the vacuum scenario, one obtains $a(\chi)\propto g^{-1/3}$. The evolution of the {elementary cell} in this case, as well as {its} comparison with the classical evolution is {also} shown {in} Fig. \ref{fig:Voffields}.

\section{LQC corrections in Einstein frame} \label{sec:einstein}

The non-{minimal} coupling between a scalar field and the Ricci scalar may be replaced by the minimally coupled system with redefined metric tensor, which leads to the GR form of the action. {This approach (the so-called Einstein frame) is equivalent to the Jordan frame analysis at the classical level. It is often more convenient to perform calculations in Einstein frame and (under the assumption, that the Jordan frame is the physical one) to express results in terms of physical variables.} Let us define $\tilde{g}_{\mu\nu}$ of the form of
\begin{equation}
\tilde{g}_{\mu\nu} = \varphi g_{\mu\nu}  \ .
\end{equation}
{In the FRW model we let}
\begin{equation}
d\tilde{t} = \sqrt{\varphi} dt\ ,\qquad \tilde{a} = \sqrt{\varphi}a\ ,
\end{equation}
{where $t$ and $a$ are cosmological time and scale factor in Jordan frame. Then the action \ref{action} in the cosmological model may be expressed as}
\begin{equation}
\tilde{S} = \int d\tilde{t}\left(-3\tilde{a}'^2\tilde{a} + \tilde{a}^3\frac{\beta}{4}\left(\frac{\varphi'}{\varphi}\right)^2-\tilde{a}^3\frac{V}{\varphi^2}\right) + S_M(\tilde{g}_{\mu\nu},\varphi) \ ,\label{eq:actionEin}
\end{equation}
where {$'$ denotes the derivative with respect to $\tilde{t}$}. The action of matter fields depends on $\varphi$ due to the transformation to Einstein frame. The action (\ref{eq:actionEin}) may be simplified with a new scalar field defined by
\begin{equation}
\phi  = \sqrt{\frac{\beta}{2}}\ln \left( \varphi\right)\ \Rightarrow\  \varphi = \exp\left(\sqrt{\frac{2}{\beta}}\phi \right) \ .
\end{equation}
Then the kinetic term in the action (\ref{eq:actionEin}) takes the canonical (i.e. GR) form and the Lagrangian looks as follows:
\begin{equation}
\mathcal{L}=-3\tilde{a}'^2\tilde{a} + \tilde{a}^3\frac{1}{2}\phi'^2-\tilde{a}^3\tilde{V}(\phi) + \mathcal{L}_M\ ,\label{LagEin}
\end{equation}
where $\tilde{V}(\phi)=\frac{V}{\varphi^2}$ ({with $\varphi$ taken as a function of $\phi$}) and $\mathcal{L}_M$ is the Lagrangian of matter fields in Einstein frame. {Similar to Eq. (\ref{eq:canonicalvariables}), the coefficient $\tilde{c}$ of the connection and the coefficient $\tilde{p}$ of the densitized triad in the FRW model can be also isolated by the symmetric reduction. Since the connection and densitized triad come from the Einstein frame, one gets their relation to those of Jordan frame as}
\begin{equation}
|\tilde{p}|=\tilde{a}^2V_o^{2/3}={\varphi |p|}\ ,\qquad\tilde{c}=\gamma\tilde{a}' = {\gamma (\dot{a}+a\frac{\dot{\varphi}}{2\varphi})=c + \gamma a \frac{\dot{\varphi}}{2\varphi}}\ ,
\end{equation}
{with the Poisson bracket $\{\tilde{c},\tilde{p}\} = \frac{\gamma}{3}$. The Lagrangian density (\ref{LagEin}) also implies $\pi_\phi=\tilde{a}^3 \phi '$}. Under the assumption that the only matter field is a scalar field $\chi$ with potential $W(\chi)$, one obtains following classical Hamiltonian in terms of {new} variables {as}
\begin{equation}
\tilde{\mathcal{H}} = \frac{\pi _{\phi }^2+e^{ \sqrt{\frac{2}{\beta }} \phi } \pi _{\chi }^2}{2 \tilde{p}^{3/2}}-\frac{3 \tilde{c}^2 \sqrt{\tilde{p}}}{\gamma ^2}+ \tilde{p}^{3/2} \left(\tilde{V}+e^{-2\sqrt{\frac{2}{\beta }} \phi } W\right)\ ,
\end{equation}
{where $\pi_\chi \equiv \tilde{a}^3 e^{-\sqrt{\frac{2}{\beta }} \phi } \chi '$ is the canonical momentum of $\chi$}. All classical equations of motion may be obtained from the Hamiltonian equations. This means that for a given function $\xi$ on the classical phase-space one obtains $\xi'=\{\xi,\tilde{\mathcal{H}}\}$.
\\*

The main motivation to implement LQC corrections in Einstein frame is that the issue of the physical interpretation of both frames is {still} open. {We need to} know how to distinguish on the experimental level between Jordan and Einstein frames LQC quantisation. {This} is a strong suggestion to analyse and compare both of them. {In this} paper {we follow the assumption that the} {Jordan} {frame is the physical one.} {So,} {to compare the quantisation in both frames we shall express the results of the Einstein frame quantisation as a function of Jordan frame variables and fields. One can still treat the Jordan frame as an underlying frame for quantisation of all degrees of freedom, from which the evolution in physical (Jordan) frame emerges.} The other reason {to consider the Einstein frame quantisation} is that the Einstein frame Hamiltonian obtains its canonical form, which {is easy} to use methods of LQC quantisation discussed in details in literature (e.g. in \cite{Bojowald:2006da}). {The procedure of the LQC quantisation of the Brans-Dicke theory in Einstein frame is similar to that in Jordan frame. But now the kinematical Hilbert space is defined over the Bohr compactification of the configuration space of $\tilde{c}$. The momentum operator $\tilde{p}$ acts on its orthonormal eigenstate $\ket{\tilde{\mu}}$ in the same way as Eq. (\ref{eq:ponmu}). In the construction of the Hamiltonian operator, one employs the holonomy $\exp(i\tilde{\bar{\mu}}\tilde{c}/2)$ with $\tilde{\bar{\mu}}=\sqrt{\Delta/|\tilde{p}|}$.} {Again, we limit ourselves to the LQC holonomy correction. Thus}, {in the} {semi-classical} {regime of} {LQC quantisation one shall transform} $\tilde{c}$ into $\sqrt{|\tilde{p}|/\Delta}\sin(\tilde{c}\sqrt{\Delta/|\tilde{p}|})${, while $\tilde{p}$ remains unchanged.} \cite{Artymowski:2012is,Bojowald:2006bz,Bojowald:2006hd}. This gives the effective Friedmann equation and equation of motion of the form of
\begin{equation}
3\tilde{H}^2 = \tilde{\rho}\left(1-\frac{\tilde{\rho}}{\rho_{\text{cr}}}\right)\ ,\qquad \frac{d^2\phi}{d\tilde{t}^2} + 3\tilde{H}\frac{d\phi}{d\tilde{t}} + \frac{dV}{d\phi} = e^{-2 \sqrt{\frac{2}{\beta }} \phi } \sqrt{\frac{1}{2\beta }} \left(4W-e^{\sqrt{\frac{2}{\beta }} \phi } \chi '^2\right)\ ,\label{eq:FriedEinLQC}
\end{equation}
where $\tilde{H} = \frac{\tilde{a}'}{\tilde{a}}$ and $\tilde{\rho} = \frac{\rho_e}{\varphi^3}$. The effective Friedmann equation may be rewritten in terms of Jordan frame metric {as}
\begin{equation}
3\left(H+\frac{\dot{\varphi}}{2\varphi}\right)^2 = \frac{\rho_e}{\varphi^2}\left(1-\frac{\rho_e}{\varphi^3\rho_{\text{cr}}}\right)\ .\label{eq:FriedEin}
\end{equation}
{Thus}, equations (\ref{eq:FriedBDJordan}) and (\ref{eq:FriedEin}) {give sufficiently different forms of the Friedmann equation. However, they both} obtain the same classical limit for $\rho_e\ll\rho_{\text{cr}}$, which is described by Eq. (\ref{eq:FriedBD}). Note that the Einstein frame quantisation does not change {the effective} equations of motion of matter fields as well as of the Brans-Dicke field. It seems natural, because the LQC quantisation is performed for the scalar field ${\phi}$ {minimally coupled to the metric $\tilde{g}_{\mu\nu}$}. This fact differs the LQC quantisation in the two different frames, since for the Jordan frame any potential term implies the LQC correction to the equation of motion of $\varphi$. {For the Einstein frame quantisation} the bounce in Einstein frame, defined by $\tilde{H}=0$, appears for $\tilde{\rho} = \rho_{\text{cr}}$, while the bounce in Jordan frame ($H=0$) requires
\begin{equation}
\frac{4\rho_e^2}{\varphi_{cr}^3\left(4\rho_e-3\dot{\varphi}_{cr}^2\right)} = \rho_{\text{cr}}\ ,\label{eq:bounceEin}
\end{equation}
{where the index $_{cr}$ denotes the value of the field at the moment of the bounce in Jordan frame.} In general, scales of bounces which originate from quantisation in Jordan frame (\ref{eq:FriedBDJordan}) and Einstein frame (\ref{eq:bounceEin}) respectively may be sufficiently different, since $\varphi_{cr}^3(1-3\dot{\varphi}_{cr}^2/4\rho_e)$ does not need to be equal to 1. Thus, not only the exact evolution, but even the scales of the bounces (understood as the value of the effective energy density at the moment, in which $H=0$) are different in the different frame quantization.
\\*

{An interesting scenario of Einstein frame quantization is the vacuum case with $\beta=3$ and non-zero potential. Then at the moment of the Jordan frame bounce one obtains $\rho_e=\varphi_{cr}^2\sqrt{\rho_{\text{cr}}V(\varphi_{cr})}$. If the $\varphi_{cr}$ is close to the minimum of the potential $V$ (which in the most realistic scenario would be in $\varphi=1$) one obtains very low scale of a Jordan frame bounce. In particular, the scale of the bounce may be close to inflationary scale and the LQC effects may be visible for the biggest scales of the power spectrum of primordial curvature perturbations.}

\section{Exact solutions of effective loop quantum Brans-Dicke cosmology in Einstein frame} \label{sec:einsteinexact}

\subsection*{Vacuum solution}

The LQC quantisation in Einstein frame provides several analytical solutions {in semi-classical regime}. For instance, the case $\rho_M = P_M = V = 0$ corresponds to the domination of a massless scalar field in the universe with {the} scale factor $\tilde{a}$. Then Eq. (\ref{eq:FriedEinLQC}) have the exact solution of the form of \footnote{More general solution of Eq. (\ref{eq:FriedEinLQC}) is {given} in Ref. \cite{Artymowski:2008sc}.}
\begin{equation}
\frac{\varphi'}{\varphi} = \phi' \propto \tilde{a}^{-3}\ \Rightarrow\ \tilde{a} = \tilde{a}_{\text{cr}}\left(1+3\rho_{\text{cr}}\tilde{t}^2\right)^{1/6}\ , \label{eq:solEin}
\end{equation}
where $\tilde{a}_{\text{cr}}$ is a value of $\tilde{a}$ at the moment of the bounce in Einstein frame. For this scenario the evolution of the Hubble parameters for the Brans-Dicke theory with LQC corrections in Jordan and Einstein frames {are} presented {respectively in} {Fig. \ref{fig:einVSjord}}
\\*

Let us note that the analytical solution in the vacuum case with $V(\varphi)=0$ may be also obtained as a function of the Brans-Dicke field. From Eq. (\ref{eq:motionBD}) and (\ref{eq:FriedEin}) one finds
\begin{equation}
\ddot{\varphi }+\frac{3\dot{\varphi }^2}{2\varphi }\left(\sqrt{\frac{\beta }{3} \left( 1-\frac{\beta  \dot{\varphi }^2}{4\varphi ^3\rho _{\text{cr}}}\right)}-1\right)=0\ \Rightarrow\ j'+\frac{3j}{2\varphi }\left(\sqrt{\frac{\beta }{3} \left( 1-\frac{\beta  j^2}{4\varphi ^3\rho _{\text{cr}}}\right)}-1\right)=0\ ,
\end{equation}
where $j=j(\varphi)=\dot{\varphi}$ and $j'=\frac{dj}{d\varphi}$. The solution of this equation is {the} following,
\begin{equation}
j=4\sqrt{\frac{\rho _{\text{cr}}}{\beta }}\frac{\varphi ^{3/2}\left(\varphi \left/\tilde{\varphi} _{\text{cr}}\right.\right)^{\frac{\sqrt{3\beta }}{2}}}{1+\left(\varphi \left/\tilde{\varphi} _{\text{cr}}\right.\right)^{\sqrt{3\beta }}}\ ,\qquad H=-\frac{1}{3}j'\ ,\qquad {a(\varphi)\propto a^{-1/3}}\ ,\label{eq:j}
\end{equation}
where $\tilde{\varphi}_{\text{cr}}\neq\varphi_{\text{cr}}$ is the value of $\varphi$ at the moment of a bounce in Einstein frame. The bounce in Jordan frame happens for $\varphi=\varphi_{\text{cr}}=\tilde{\varphi }_{\text{cr}}\left(\frac{\sqrt{\beta} +\sqrt{3}}{\sqrt{\beta} -\sqrt{3}}\right)^{1/\sqrt{3\beta }}$ and for energy density $\tilde{\rho}(\varphi_{\text{cr}})=\left(1-3\left/\beta\right.\right) \rho _{\text{cr}}$. In the limit of $\beta\gg 1$ one obtains $\tilde{\varphi}_{\text{cr}}\simeq\varphi_{\text{cr}}$ and $\tilde{\rho}(\varphi_{\text{cr}})\simeq\rho_{\text{cr}}$.

\subsection*{The $V(\varphi)=\Lambda\varphi^2$ case}

The other analytical solution comes from the case $V(\varphi)=\Lambda\varphi^2$ {and} $\rho_M=0$. Then {in Einstein frame one obtains} $\tilde{V} = \Lambda$, so {if the $\varphi$} field has {a square potential in Jordan frame, it} gives the universe filled with massless scalar field $\phi$ and cosmological constant $\Lambda$ {in Einstein frame}. Let us assume that at the moment of a bounce in Einstein frame, when $\tilde{H} = 0$, one obtains $\Lambda=(1-\alpha)\rho_{\text{cr}}$ and $\frac{1}{2}\dot{\phi}_{\text{cr}}^2 = \alpha\rho_{\text{cr}}$. This comes from the fact that $\Lambda+\frac{1}{2}\dot{\phi}_{\text{cr}}^2 = \rho_{\text{cr}}$. The $\alpha$ parameter has an interpretation of percentage contribution of $\rho_m$ to the critical energy density. Usually one expects  a cosmological constant to be subdominant around the big bounce. {Thus} it is natural to consider $|1-\alpha|\ll 1$. One shall note that $\alpha<1$ and $\alpha>1$ corresponds to $\Lambda>0$ and $\Lambda<0$ respectively. The exact solution of Eq. (\ref{eq:FriedEinLQC}) looks as follows:
\begin{eqnarray}
\tilde{a}=\tilde{a}_{\text{cr}}\left(\frac{\cosh\left(6\sqrt{\theta}\tilde{t}\right)-2\alpha+1}{2(1-\alpha)}\right)^{\frac{1}{6}}\qquad \text{for}\qquad \Lambda>0\ , \label{eq:a(t)L>0LQC}\\
\tilde{a}=\tilde{a}_{\text{cr}}\left(\frac{-\cos\left(6\sqrt{\theta}\tilde{t}\right)+2\alpha-1}{2(\alpha-1)}\right)^{\frac{1}{6}}\qquad \text{for}\qquad \Lambda<0 \ ,\label{eq:a(t)L<0LQC}
\end{eqnarray}
where $\theta=\sqrt{|1-\alpha|\alpha\rho_{\text{cr}}/3}$. It is easy to show that for $\alpha\leq 1/2$ one obtains $\frac{d}{d\tilde{t}}\tilde{H}>0$ at any time. In such a case $\tilde{H}$ grows to its finite maximal value $\tilde{H}_{max}=\sqrt{\theta}$. {For $\alpha>1/2$} $\tilde{H}$ grows initially to reach it's global maximum $\tilde{H}_{max}=\sqrt{\rho_{\text{cr}}/12}$ {at} $\rho=\rho_{\text{cr}}/2$. Later on $\tilde{H}$ decreases together with $\tilde{\rho}$. For $\sqrt{\theta}t\ll 1$ one finds $\cosh(6\sqrt{\theta}\tilde{t})\simeq 1+6\alpha\rho_{\text{cr}}(1-\alpha)\tilde{t}^2$ and $\cos(6\sqrt{\theta}\tilde{t})\simeq 1-6\alpha\rho_{\text{cr}}(1-\alpha)\tilde{t}^2$. Thus, for both positive and negative cosmological constant, around a bounce one recovers the solution (\ref{eq:solEin}). Let us note that the case of $V = \Lambda\varphi^2$ is possible to solve analytically only for the quantisation in Einstein frame. {An} interesting feature of this model is that, although $\varphi$ has a potential term, one may still use it as a time variable, because $\dot{\varphi}>0$ ia always valid. However, this is the case only in the classical limit or for the LQC quantisation in Einstein frame. For $\varphi\gg 1$ the considered potential is a good approximation of the potential of the Starobinsky inflation \cite{Starobinsky:1980te}, for which $V\propto(\varphi-1)^2$. However, $V=\Lambda\varphi^2$ does not provide the graceful exit, and it generates too flat power spectrum of initial curvature perturbations.

{\subsection*{With massless scalar field}}

\begin{figure}[h]
\centering
\includegraphics*[height=5.5cm]{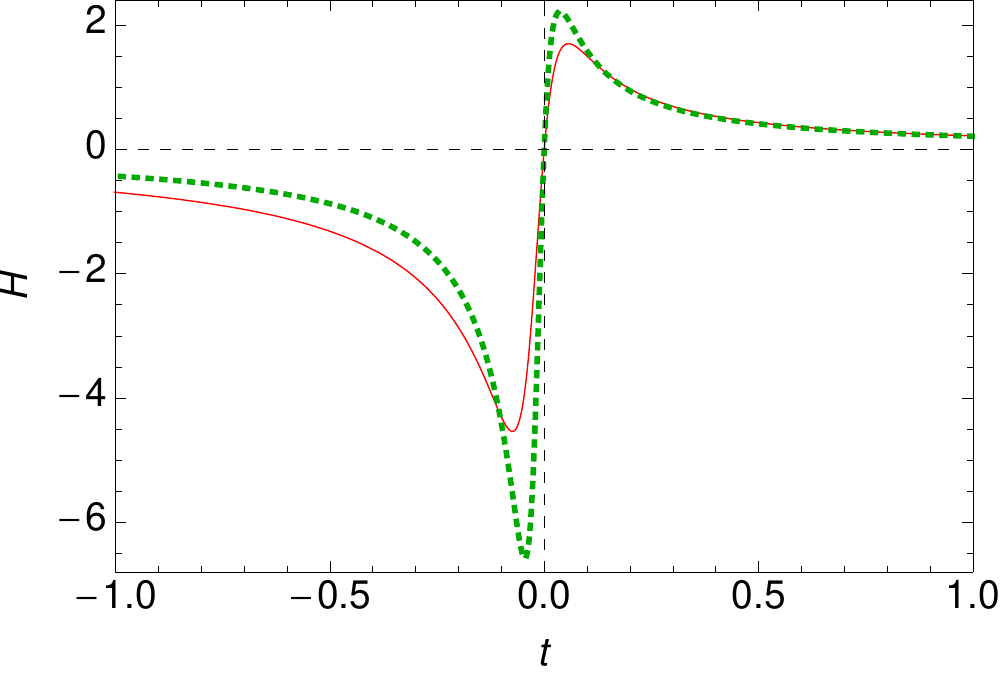}
\hspace{0.8cm}
\includegraphics*[height=5.5cm]{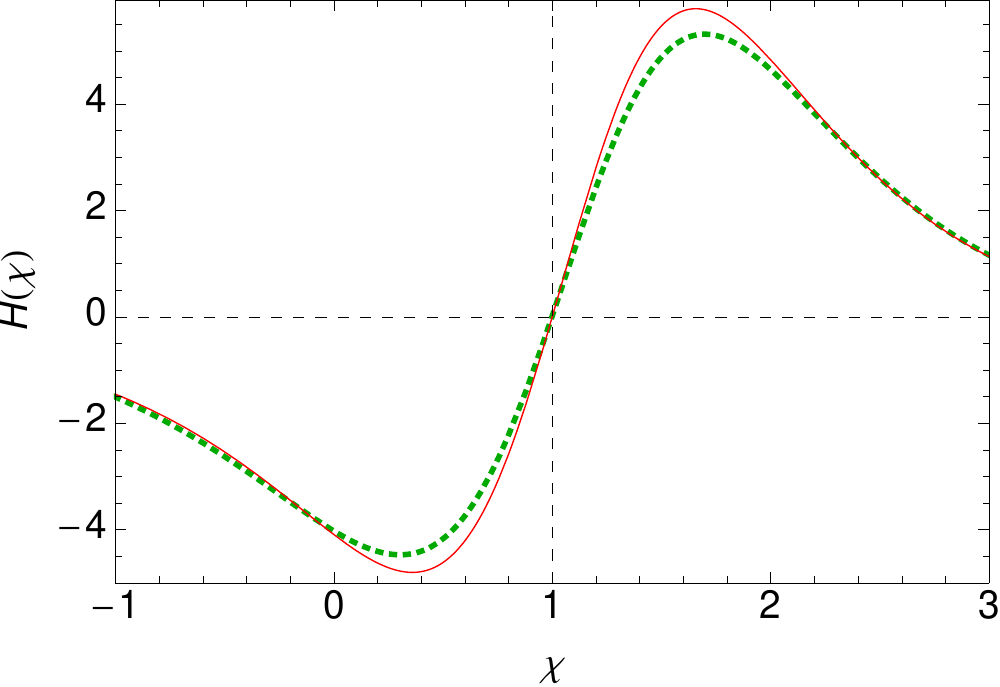}
\caption{\it All panels lines present the analytical solutions for the physical Hubble parameter $H$ as a function of cosmological time $t$ (left panel, vacuum solution, $\beta=5$, $\varphi_{\text{cr}}=1$) or $\chi$ (right panel, massless scalar field solution, $\beta=15$, $\chi_{\text{cr}}=1$, $B=0$, $C=1$) for the LQC quantisation in Jordan (red line) and Einstein (dashed green line) frames. The $t=0$ or $\chi=1$ corresponds to the moment of the bounce in Jordan frame. For $t>t_{pl}$ both methods of quantisation {approach} the same {evolution trajectory}. The comparison takes place in Jordan frame due to our assumption that this frame is the physical one.}
\label{fig:einVSjord}
\end{figure}

For the Einstein frame quantisation one may also obtain analytical solutions {of the effective theory} in the case of $V=W=0$. Since classical equations of motion of matter fields are still valid for the {effective theory of the} Einstein {frame} quantisation, one may use the relation (\ref{eq:phiofvarphi}) and the semi-classical Friedmann equation (\ref{eq:FriedEinLQC}) to obtain equation of motion of only one degree of freedom, which is $\chi$. Again, we define $h(\chi)=\dot{\chi}$, which gives following equation of motion,
\begin{equation}
\frac{d h}{d\chi}-\frac{3(B \beta -\chi )h}{2 C \beta +(2 B \beta -\chi ) \chi }+\sqrt{\frac{3}{2\beta }(2 C \beta +(2 B \beta -\chi ) \chi ) } \sqrt{\tilde{\rho } \left(1-\frac{\tilde{\rho }}{\rho _{\text{cr}}}\right)}\ ,
\end{equation}
where
\begin{equation}
\tilde{\rho} = \frac{2 \beta ^3 \left(2 C+B^2 \beta \right) h^2}{\left(2 C \beta +2 B \beta  \chi -\chi ^2\right)^3}\ .
\end{equation}
This equation has following exact solution,
\begin{equation}
h=\sqrt{\frac{2\rho _{\text{cr}}}{\beta ^3 \left(2 C+B^2 \beta \right)}}\frac{\left(\frac{B \beta +\sqrt{\beta } \sqrt{2 C+B^2 \beta }-\chi }{B \beta +\sqrt{\beta } \sqrt{2 C+B^2 \beta }-\tilde{\chi} _{\text{cr}}}\right)^{ \frac{1}{2}\sqrt{3\beta }}\left(\frac{-B \beta +\sqrt{\beta } \sqrt{2 C+B^2 \beta }+\chi }{-B \beta +\sqrt{\beta } \sqrt{2 C+B^2 \beta}+\tilde{\chi}_{\text{cr}}}\right)^{\frac{1}{2}\sqrt{3\beta}}}{\left(\frac{B \beta +\sqrt{\beta } \sqrt{2 C+B^2 \beta }-\chi }{B \beta +\sqrt{\beta } \sqrt{2 C+B^2 \beta }-\tilde{\chi} _{\text{cr}}}\right)^{\sqrt{3\beta }}+\left(\frac{-B \beta +\sqrt{\beta } \sqrt{2 C+B^2 \beta }+\chi }{-B \beta +\sqrt{\beta } \sqrt{2 C+B^2 \beta }+\tilde{\chi} _{\text{cr}}}\right)^{\sqrt{3\beta }}}(2 C \beta +(2 B \beta -\chi ) \chi )^{3/2}
\end{equation}
where $\tilde{\chi}_{\text{cr}}\neq\chi_{\text{cr}}$ represent the moment in which $\tilde{H}=0$. Similarly to the Jordan frame quantisation, the Hubble {parameter} in Jordan frame is of the form of $H(\chi)=-\frac{1}{3}\frac{d h}{d\chi}$. This result is {also} compared with the Jordan frame quantisation {in} Fig. \ref{fig:einVSjord}.

\section{Conclusions}\label{sec:concl}

In this paper we have considered the issue of the semi-classical evolution of {the universe} in {both Jordan and Einstein frames} {of} LQC Brans-Dicke theory with scalar potential $V(\varphi)$ {coupled} with {an} additional scalar field $\chi$ with a potential $W(\chi)$. {The Hamiltonian formalism of the corresponding cosmological models are presented in terms of geometrical variables as well as connection variables respectively.}
\\*

{To compare the Jordan frame with Einstein frame of Brans-Dicke theory, the same cosmological model is quantized by LQC method in both frames separately. We then consider the effective equations with LQC holonomy corrections resulted from the different frames quantisation. In particular, the effective Friedmann equations and equations of motion for the scalar fields are obtained in both frames. In the Jordan frame quantisation it is} shown that not only the potential term of {the} Brans-Dicke field {$\varphi$}, but also $W(\chi)$ {can lead to corrections to the effective} equation of motion of $\varphi$. In the $\rho_e\ll\rho_{\text{cr}}$ limit the classical Brans-Dicke theory {can be} recovered {from the effective theory}. In the $W=0$ case equations of motion for scalar fields and the semi-classical Friedmann equation are of the the same form as for the Brans-Dicke theory with and without a potential.
\\*

{Analytical solutions have been found for the effective equations resulted from both frames quantisation without potential in the vacuum case and in the additional massless scalar field case separately. In the vacuum case, the Brans-Dicke scalar field can be employed as an internal time. In the matter coupled case, the matter scalar field $\chi$ is used as a time variable, and the analytical relation} {(valid in the effective theory of both frames and in the classical theory)} {between the two scalar fields is obtained. In all those solutions the bouncing evolution of the scale factor has been obtained around Planck regime. However, The quantisation of different frames lead to different scales of the bounces} {of the scale factor $a(t)$. }{The Jordan frame quantisation and the Einstein frame quantisation require} $\rho_e=\rho_{\text{cr}}$ and $\frac{4\rho_e^2}{\varphi_{cr}^3\left(4\rho_e-3\dot{\varphi}_{cr}^2\right)} = \rho_{\text{cr}}$ respectively {for the occurrence of the bounce. Hence, different frame quantisation gives different physics. Moreover, as shown in Fig. \ref{fig:Voffields}, the evolutional trajectories of the Jordan frame volume of the elementary cell of the universe  are different for different frames quantisation.}
\\*

{However, as shown in Fig. \ref{fig:einVSjord} in the vacuum case the difference of the evolutional trajectories of the Jordan frame Hubble parameter disappears for time $t>t_{pl}$ after the bounces.} {Therefore, it may be difficult to distinguish between frames quantisation by observations of primordial inhomogeneities. {The situation may change by} the fine-tuning of the initial conditions of the Einstein frame quantisation, which (comparing to the Jordan frame quantisation) may sufficiently decrease the scale of the Jordan frame bounce. In particular this could lead to the Jordan frame bounce around the GUT scale with the period of superinflation visible in the power spectrum of primordial inhomogeneities. This issue, analyzed in the context of Starobinsky inflation, shall be the goal of our future work.}
\\*

{Another interesting feature of the bounce in loop quantum Brans-Dicke cosmology, explored in Fig. \ref{fig:Voffields}, is that, unlike the case of LQC of GR, the evolutional trajectories of the two sides of the bounce point are obviously asymmetric. As a by-product, we also find an analytical solution with Jordan frame potential $V=\Lambda\varphi^2$ in the theory of Einstein frame quantisation. }

\begin{acknowledgements}
This work is supported by NSFC (No.11235003). and
the Fundamental Research Funds for the Central University of China
under Grant No.2013ZM107. M.A. would also like to acknowledge China Postdoctoral
Science Foundation for financial support.

\end{acknowledgements}

\end{document}